\documentstyle[12pt]{article}
\newcommand{\sq}{\tilde{q}}
\newcommand{\ca}{c_\alpha}
\newcommand{\sa}{s_\alpha}
\newcommand{\sbe}{s_\beta}
\newcommand{\cb}{c_\beta}

\newcommand{\std}{s_W^2}
\newcommand{\sab}{s_{\alpha+\beta}}
\newcommand{\cab}{c_{\alpha+\beta}}
\newcommand{\beq}{\begin{equation}}
\newcommand{\eeq}{\end{equation}}
\newcommand{\beqn}{\begin{eqnarray}}
\newcommand{\eeqn}{\end{eqnarray}}
\newcommand{\stackM}{\stackrel{\scriptstyle >}{{ }_{\sim}}}

\begin{document}

\thispagestyle{empty}
\def\pubnum{377}
\def\data{November, 1995}

\begin{flushright}
{\parbox{3.5cm}{
UAB-FT-377

November, 1995

hep-ph/9511402
}}
\end{flushright}
\vspace{3cm}
\begin{center}
\begin{large}
\begin{bf}
STRONG EFFECTS ON THE HADRONIC WIDTHS OF THE NEUTRAL HIGGS BOSONS
IN THE MSSM
\\
\end{bf}
\end{large}
\vspace{1cm}
J.A. COARASA, Ricardo A. JIM\'ENEZ, Joan SOL\`A\\

\vspace{0.25cm}
Grup de F\'{\i}sica Te\`orica\\
and\\
Institut de F\'\i sica d'Altes Energies\\
\vspace{0.25cm}
Universitat Aut\`onoma de Barcelona\\
08193 Bellaterra (Barcelona), Catalonia, Spain\\
\end{center}
\vspace{0.3cm}
\hyphenation{super-symme-tric}
\hyphenation{com-pe-ti-ti-ve}
\begin{center}
{\bf ABSTRACT}
\end{center}
\begin{quotation}
\noindent
\hyphenation{ob-ser-va-bles}
\noindent
We analyze the correlation of QCD one-loop effects on the
partial widths of the three neutral Higgs bosons of the MSSM decaying into
quark-antiquark pairs. The SUSY-QCD effects turn out to be
comparable or even larger
than the standard QCD effects and are slowly decoupling
in a wide window of the parameter space.
Our study is aimed at elucidating the possible supersymmetric nature of
the neutral Higgs bosons that might be discovered in the near future at the
Tevatron and/or at the LHC. In particular, we point out the presence
of potentially large SUSY corrections to the various neutral
Higgs production cross-sections.
\end{quotation}

\baselineskip=6.5mm  %(FOR PREPRINT)

\newpage

To glimpse into the relevance of addressing the issue of the
width of a Higgs boson, notice that
if a heavy neutral Higgs is discovered and is found to have
a narrow width, it would certainly not be the SM Higgs, whilst
 it could be a SUSY Higgs.
In fact, a heavy enough SM Higgs boson is expected to rapidly
develop a broad width through decays
into gauge boson pairs whereas the SUSY Higgs bosons cannot
in general be that broad since their couplings to gauge bosons
are well-known to be suppressed\,\cite{Hunter}.
In compensation, their couplings to
fermions (especially to heavy quarks)
can be considerably augmented.
Thus the width of a
SUSY Higgs should to a great extent be given by its hadronic width;
even so a heavy $H^0$ and $A^0$ is in general
narrower than a SM Higgs of the same mass.
Alternatively, if the discovered
neutral Higgs is sufficiently light that it cannot decay into gauge boson
pairs, its decay width into relatively heavy fermion pairs such as
$\tau^+\,\tau^-$, and especially into $b\,\bar{b}$, could be much larger than
that of the SM Higgs, because of
$\tan\beta$-enhancement of the fermion couplings\,\cite{Hunter}.
Hence, it becomes clear that the hadronic width may play a very
important role in the study of the MSSM higgses, already at the tree-level.

As is well-known\,\cite{Hunter}, a generic two-doublet Higgs sector is
composed of a charged Higgs pseudoscalar boson, $H^{\pm}$,
a neutral CP-odd (``pseudoscalar'') boson, $A^0$, and two neutral
CP-even Higgs ``scalars'', $h^0$ and $H^0$ ($M_{h^0}< M_{H^0}$).
In the specific case of the MSSM, and
because of the supersymmetric constraints,
only two parameters, e.g. $(M_{A^0},\tan\beta$), are independent at
the tree-level\,\cite{Hunter}. Therefore, in the MSSM
definite predictions can be made and tested.

The aim of this
work is to complete the analysis of the strong SUSY corrections to the
hadronic decay widths of the Higgs bosons of the MSSM that we have initiated in
the
companion paper\,\cite{Ricard}. In the latter reference whose notation and
definitions we assume hereafter we have defined
two interesting domains of the general MSSM parameter,
the so-called Regions I and II, where the physics
of the supersymmetric Higgs bosons can be especially relevant.
Depending on the region of parameter space considered, not all the Higgs
particles
of the MSSM are allowed to decay hadronically in a significant manner.
On the one hand, the
process $H^+\rightarrow t\,\bar{b}$, which requires $M_{H^+}>m_t+m_b\sim
180\,GeV$,
is permitted in Region II
and the SUSY-QCD corrections can be relevant in that region\,\cite{Ricard}.
When $H^+\rightarrow t\,\bar{b}$ is allowed, such a decay is by far the main
hadronic
decay mode of a SUSY charged Higgs boson.
If, however, the condition $M_{H^+}>m_t+m_b$
is not satisfied, the remaining hadronic decays available to $H^+$ are not so
appealing since the corresponding branching fractions lie below
the leptonic $\tau$-mode
$H^+\rightarrow \tau^+\,{\nu}_{\tau}$.
On the other hand,
we may turn our attention to the various hadronic neutral Higgs decays
$\Phi^i\rightarrow q\bar{q}$
($\Phi^1\equiv A^0,\,\Phi^2\equiv h^0,\,\Phi^3\equiv H^0$).
Of these, we will neglect the decays leading to
light $q\,\bar{q}$ final states since their branching ratio is very small.
Thus, for the lightest neutral Higgs, $h^0$, we will concentrate on just
the decay $h^0\rightarrow b\,\bar{b}$,
whereas for $A^0, H^0$ (which can be arbitrarily heavy) we shall consider the
channels $A^0, H^0\rightarrow b\,\bar{b}$ and $A^0, H^0\rightarrow t\,\bar{t}$.

Moreover, should the physical domain of the MSSM
parameter space turn out to lie in Regions I or II, then we shall see that
the hadronic widths of the MSSM Higgs bosons must incorporate
important virtual SUSY signatures.
The latters could be extracted from measured quantities by subtracting
the corresponding conventional QCD effects, which can
be easily computed by adapting the results of Refs.\cite{Braaten,Drees}.
Knowledge of the SUSY corrections could be determinant to trace the
nature of those scalars and establish whether they are
truly supersymmetric Higgs particles.

Although we elaborate here mostly on the Higgs
strategies at hadron colliders, such as the Tevatron and especially the LHC,
it should be clear that the kind of effects that we wish to study have an
impact
on Higgs physics in $e^+\,e^-$ machines as well, where the neutral Higgs states
can
be produced through e.g.\,
$e^+\,e^-\rightarrow Z\,h^0(H^0)$ and
$e^+\,e^-\rightarrow A\,h^0(H^0)$. The observed cross-sections for these
processes
are equal to the production cross-sections times the Higgs
branching ratios.  Thus,
in an $e^+\,e^-$ environment one aims more at
a measurement of the various
branching ratios (or, more precisely: ratios of branching ratios)
of the fermionic Higgs decay modes rather than of the
partial widths themselves.
For instance, in $e^+\,e^-$ one would naturally address the measurement of
$BR(\Phi^i\rightarrow b\,\bar{b})/BR(\Phi^i\rightarrow \tau^+\,\tau^-)$;
in fact, this observable should receive large SUSY-QCD corrections
if $\Phi^i\rightarrow b\,\bar{b}$ proves to be, as we shall see,
very sensitive to the strong supersymmetric effects.

In hadron machines an actual measurement
of the hadronic partial widths and in general of the effective hadronic
vertices
$\Phi^i\,q\,\bar{q}\, (q=t,b)$ should be feasible.
Let us briefly remind of the five basic
mechanisms for neutral
Higgs production in a hadron collider\,\cite{Denegri}.
They have been primarily described for
the SM Higgs, $H^0_{SM}$, but can be straightforwardly extended
to the three neutral higgses, $\Phi^i$, of
any two-doublet Higgs sector (see Fig.1a. for a sketch of some of these
mechanisms):
\begin{itemize}
\item{(i)}
Gluon-gluon fusion: $g\,g\rightarrow \Phi^i$;
\item{(ii)} $WW(ZZ)$ fusion:
$q\,q\rightarrow q\,q\,\Phi^i$;
\item{(iii)}
 Associated $W(Z)$ production: $q\,\bar{q}\rightarrow W (Z)\,\Phi^i$;
\item{(iv)} $t\,\bar{t}$ fusion:
 $g\,g\rightarrow t\,\bar{t}\,\Phi^i$, and
\item{(v)} $b\bar{b}$ fusion:
 $g\,g\rightarrow b\,\bar{b}\,\Phi^i$.
\end{itemize}
It has been known for a long time\,\cite{Georgi} that in the SM, where only
one neutral Higgs $H^0_{SM}$
is present, mechanism (i) provides the
dominant contribution over most of the accessible range.
For very large (obese) SM Higgs mass, however, mechanism (ii) eventually
takes over; the rest of the mechanisms are subleading, and in particular
$b\,\bar{b}$ fusion is negligible in the SM.

Remarkably enough, this situation could drastically change
in the MSSM. As noted
above for the Higgs decays, also the production mechanisms of the
MSSM Higgs scalars can be very different from the SM\,\,\cite{Denegri}.
For instance, whereas
one-loop $g\,g$-fusion in the SM is dominated by a top quark in the
loop, this is not always so in the MSSM where the new couplings
turn out to enhance, at high $\tan\beta$,
the $b$-quark loops and make them fully competitive with the top quark loops
 (Fig. 1a). Furthermore, mechanism (ii) becomes
suppressed by the SUSY couplings; e.g.
in Region I the lightest
neutral Higgs, $h^0$, has a very small coupling to the weak gauge bosons
as compared to $H^0_{SM}$. In this respect
the situation with the CP-odd scalar, $A^0$,
is even worse, for it can never
be produced by mechanism (ii) at the tree-level.
In contrast, $b\,\bar{b}$ fusion (Fig.1a), which was negligible in the SM,
can be very important in the MSSM at large
$\tan\beta$, especially in Region I but also in Region II.
As a matter of fact, for
large enough $\tan\beta$,
the $b\,\bar{b}$-fusion cross-section can be larger than
that for any mechanism for producing a SM Higgs boson
of similar mass\,\cite{Denegri}.

Our interest in the production mechanisms mentioned above stems from the
fact that the radiative effects could play a crucial role.
This is true already in the SM.
For example, the conventional QCD corrections to $g\,g\rightarrow H^0_{SM}$,
which is the dominant process
for the production of a light and an intermediately heavy Higgs boson,
are known to be large\,\cite{DSZ}.
A similar conclusion holds for an obese SM Higgs boson
produced at very high energies by means of the
$WW(ZZ)$-fusion mechanisms; here, again,
non-negligible radiative effects do appear
\,\cite{Marciano}. Therefore, the production cross-section
for $H^0_{SM}$ is expected to acquire valuable quantum corrections
both for light and  for heavy Higgs masses. This is not so, however,
for the corresponding width.
In fact, only for a heavy SM Higgs, namely, with a mass above the vector boson
thresholds, the corrections to its decay width can
be of interest; for a light SM Higgs, instead,
light enough that it cannot decay into gauge boson pairs, the decay width
is very small and thus the corresponding quantum effects
are of no practical interest.

Now, in contradistinction to the SM case,
the hadronic vertices $\Phi^i\,q\,\bar{q}\, (q=t,b)$
could be the most significant interactions for MSSM higgses,
irrespective of the value of the Higgs masses. In fact, these vertices
can be greatly enhanced.
Therefore, if large radiative corrections may modify the effective structure
of these interactions, it is clear that they should be taken into account and
could be of much practical interest.
In what follows we shall substantiate that in the MSSM
the $\Phi^i\,b\bar{b}$ and $\Phi^i\,t\bar{t}$ vertices involved in
mechanisms (i), (iv) and (v) above could receive very large SUSY-QCD
corrections (in some cases above $50\%$) and so, if these effects are there,
they will be reflected in the Higgs boson
partial widths and to a large extent also in the production cross-sections.
In this respect
the aforementioned $b\,\bar{b}$-fusion
mechanism, which is highly operative
at large $\tan\beta$, can be very sensitive to these SUSY-QCD corrections.
To our knowledge, these matters have not been discussed
in the literature and could  play a momentous role
to decide whether a neutral Higgs hypothetically produced in a
hadron collider is supersymmetric or not.

While it goes beyond the scope of this note
to compute the SUSY-QCD corrections to the production processes
themselves, we have performed a detailed analysis of the partial
decay widths, which are the canonical observables that
should be first addressed to probe the
new quantum corrections to the
basic interaction vertices. In this way, a definite prediction is made on the
properties of a physical observable, and moreover
this should suffice both
to exhibit the relevance of the SUSY quantum effects
and to  demonstrate the necessity to incorporate these corrections
in a future, truly
comprehensive, analysis of the cross-sections, namely, an analysis where
one would include the quantum effects on all the relevant
production mechanisms within the framework
of the MSSM.

Let us now concentrate on the
diagrams depicted in Fig.1b. Since we
adopt the same framework as in the companion paper\,\cite{Ricard},
we shall obviate all the unessential details already defined
there.
The interaction Lagrangian describing the $\Phi^i\,q\,\bar{q}$-vertex
in the MSSM is:
\beq
 {\cal L}_{\Phi q q}= \frac{g m_q}{2 M_W}\Phi^i \bar{q}
                       \left[ a_L^i(q) P_L +a_R^i(q) P_R\right] q\,.
\label{eq:LHqq}
\eeq
We shall focus on top and bottom quarks ($q=t,b$).
In a condensed and self-explaining notation we have defined
\beqn
a_R^1(t,b)&=&-a_L^1(t,b)=(i\cot\beta,i\tan\beta)\,,\nonumber\\
a_R^2(t,b)&=&a_L^2(t,b)=(-\ca/\sbe,\sa/\cb)\,,\nonumber\\
a_R^3(t,b)&=&a_L^3(t,b)=(-\sa/\sbe,-\ca/\cb)\,,
\label{eq:aLaR}
\eeqn
with $\ca\equiv\cos\alpha, \sbe\equiv\sin\beta$ etc. (Angles $\alpha$ and
$\beta$
are related in the usual manner prescribed by the MSSM\,\cite{Hunter}.)
Apart from the SUSY-QCD interactions involving gluinos
and squarks, a very relevant piece of our calculation is the
interaction Lagrangian between
neutral higgses and squarks. In compact notation, it can be cast as follows:
\begin{equation}
{\cal L}_{\Phi\sq\sq}= \frac{-g}{2 M_W}\Phi^i {\sq_a}^{*}
               G_{i\ ab}^{(q)}\sq_b\,,
\label{eq:LHsqsq}
\end{equation}
where we have introduced the mass-eigenstate coupling matrices
\beq
G_{i}^{(q)}= R^{(q)\dagger}\,\hat{G}_{i}^{(q)}\,R^{(q)}\,,
\eeq
related to the corresponding
weak-eigenstate coupling matrices, $\hat{G}_{i}^{(q)}$, by means of the
rotation matrices $R^{(q)}$. The latters diagonalize the stop and
sbottom mass matrices defined in eqs.(8)-(9) of Ref.\cite{Ricard}.
For the $t\,\bar{t}$\, final states, we have
\beqn
  \hat{G}_1^{(t)}\left[\cot\beta\right]&=&
    \pmatrix{
        0  & - i m_t\left(\mu+A_t \cot\beta\right)
        \cr
        i m_t\left(\mu+A_t \cot\beta\right) & 0}\,,\nonumber\\
\nonumber\\
  \hat{G}_2^{(t)}\left[\ca,\sa,\sbe\right]&=&
     \pmatrix{
        -2 M_Z^2
        \left(T_3^{(t)}-Q^{(t)}\std\right)\sab+\frac{2 m_t^2\ca}{\sbe} &%
        \frac{m_t}{\sbe}\left(\mu\sa+A_t\ca\right)
        \cr
        \frac{m_t}{\sbe}\left(\mu\sa+A_t\ca\right)&
        -2 M_Z^2Q^{(t)}
        \std\sab+\frac{2 m_t^2\ca}{\sbe}}\,,\nonumber\\
\nonumber\\
  \hat{G}_3^{(t)}\left[\ca,\sa,\sbe\right]&=&
     \pmatrix{
        2 M_Z^2
        \left(T_3^{(t)}-Q^{(t)}\std\right)\cab+\frac{2 m_t^2\sa}{\sbe} &%
        \frac{m_t}{\sbe}\left(-\mu\ca+A_t\sa\right)
        \cr
        \frac{m_t}{\sbe}\left(-\mu\ca+A_t\sa\right)&
        2 M_Z^2 Q^{(t)}
        \std\cab+\frac{2 m_t^2\sa}{\sbe}}\,,
\label{eq:G123}
\eeqn
with $\sab\equiv\sin(\alpha+\beta)$ etc. and $Q^{(q)},T^{(q)}_3$ the
electric charge and 3rd
component of weak isospin.
For the $b\,\bar{b}$ final states, the following replacements
are to be performed with respect to the $\hat{G}_i^{(t)}[...]$
in eq.(\ref{eq:G123}):
\beqn
\hat{G}_1^{(t)}&\rightarrow&\hat{G}_1^{(b)}\left[\tan\beta\right]\,,\nonumber\\
\hat{G}_2^{(t)}&\rightarrow&\hat{G}_2^{(b)}
\left[\sa,\ca,-\cb\right]\,,\nonumber\\
\hat{G}_3^{(t)}&\rightarrow&\hat{G}_3^{(b)}\left[\sa,\ca,\cb\right]\,.
\eeqn
The one-loop renormalized vertices for any of the relevant
hadronic decays $\Phi^i\rightarrow q\,\bar{q}$\,
are derived from the on-shell renormalized Lagrangian and
can be parametrized in terms of two bare form factors
$K_L^i(q)$, $K_R^i(q)$ and the corresponding mass and wave-function
renormalization
counterterms $\delta m_q$ and
$\delta Z_{L,R}^q$ associated to the external quark lines:
\beq
  O^i(q) = \frac{ g\, m_q}{2 M_W}
        \left[ a_L^i(q)\left(1+O_L^i(q)\right) P_L +
          a_R^i(q)\left(1+O_R^i(q)\right) P_R\right]\,,
\eeq
the renormalized form factors being
\beqn
  O_L^i(q)&=&K_L^i(q)+\frac{\delta m_q}{m_q}+\frac{1}{2}\delta Z_L^q+
                                   \frac{1}{2}\delta Z_R^q\,,
  \nonumber\\
  O_R^i(q)&=&K_R^i(q)+\frac{\delta m_q}{m_q}+\frac{1}{2}\delta Z_L^q+
                                   \frac{1}{2}\delta Z_R^q\,.
\label{eq:OLOR}
\eeqn
For each $\Phi^i= A^0,\, h^0, H^0$ decaying into $q\,\bar{q}$ a
straightforward calculation of the diagrams in Fig.1b yields a generic
contribution
of the form (summation is understood over $a,b$)
\begin{eqnarray}
  K_L^i(q)&=&8\pi\alpha_s i C_F \frac{G_{i\ ab}^{(q)}}{a_L^i(q)}
      \left[ R_{1a}^{(q)}R_{1b}^{(q)*}\left(C_{11}-C_{12}\right) +
      R_{2a}^{(q)}R_{2b}^{(q)*}C_{12}+
     \frac{m_{\tilde{g}}}{m_q} R_{2a}^{(q)}R_{1b}^{(q)*}C_{0}\right]\,,
  \nonumber\\
  K_R^i(q)&=&8\pi\alpha_s i C_F \frac{G_{i\ ab}^{(q)}}{a_R^i(q)}
      \left[ R_{2a}^{(q)}R_{2b}^{(q)*}\left(C_{11}-C_{12}\right) +
      R_{1a}^{(q)}R_{1b}^{(q)*}C_{12}+
      \frac{m_{\tilde{g}}}{m_q} R_{1a}^{(q)}R_{2b}^{(q)*}C_{0}\right].
\label{eq:KLKR}
\end{eqnarray}
The explicit expressions for the mass and wave-function renormalization
counterterms
are borrowed from Ref.\cite{GJS} and will not be repeated here, and
the various $3$-point functions in eq.(\ref{eq:KLKR}) have the arguments
$C_{...}=
C_{...}(p,p^{\prime},m_{\tilde{g}},m_{\tilde{q}_a},m_{\tilde{q}_b})$
\cite{GJSH};
they carry indices $a,b$ summed over. Finally, $C_F=(N_C^2-1)/2N_C=4/3$
follows from summation over color indices.

At the end of the day, the relative SUSY-QCD correction to each decay
width  of $\Phi^i\rightarrow q\,\bar{q}$
with respect to the corresponding tree-level width reads as follows,
\beq
\delta_{\tilde{g}}^i(q)={\Gamma^i(q)-\Gamma_0^i(q)\over \Gamma_0^i(q)}=
Re [O_L^i(q)+O_R^i(q)]\,,
\label{eq:deltag}
\eeq
where $\Gamma^i(q)\equiv \Gamma (\Phi^i\rightarrow q\,\bar{q})$ is the
corrected
width, and
\beqn
& &  \Gamma_0^i(q)=\left(\frac{N_C G_F}{4
\pi\sqrt{2}}\right)\left|a_R^i(q)\right|^2\,
  M_{\Phi^i}\,m_q^2\,
  \lambda^{(1/2+j)}\,(1,\frac{m_q^2}{M^2_{\Phi^i}}
,\frac{m_q^2}{M^2_{\Phi^i}})\,,\nonumber\\
& & (j=0\ \  {\rm for}\ \  i=1\ \ {\rm  and}\ \ j=1\ \ {\rm for}\ \  i=2,3)\,,
\label{eq:treelevel}
\eeqn
are the corresponding tree-level widths.
The upshot of our exhaustive numerical analysis is synthesized in Figs.2-5.
We treat the sbottom and stop mass matrices as in Ref.\cite{Ricard}.

In Fig.2a, where we fix $M_{A^0}=60\,GeV$, we study the dependence of the
SUSY-QCD
correction (\ref{eq:deltag})
on the Higgs mixing mass parameter $\mu$ for the three decays
$\Phi^i\rightarrow b\,\bar{b}$.
We immediately gather that the sign of the correction is opposite to that of
$\mu$.
For $A^0$ and $h^0$ the correction is basically the same and can reach large
values,
e.g. $|\delta_{\tilde{g}}|\simeq 50\%$ at $\tan\beta=30$
and $|\mu|\simeq 100\,GeV$. As stressed in Ref.\cite{Ricard}, the origin of the
large SUSY-QCD contributions obtained in the presence of final states
involving the $b$-quark
can be ascribed to their self-energy renormalization effects\,\cite{SO10},
which in our case go to the counterterm $\delta m_b/m_b$ on eq.(\ref{eq:OLOR}).
We remark that the corrections affecting the $b\,\bar{b}$ final
states are larger for
$H^0$ than for $A^0$ and $h^0$. The drawback, however, is that the
huge effects obtained for $H^0\rightarrow b\,\bar{b}$  at the highest
values of $\tan\beta$
correspond to the smallest tree-level decay widths.
In contrast, the less ambitious but still quite
respectable quantum effects on  $A^0\rightarrow b\,\bar{b}$ are
larger the larger
is its decay width.

In Fig.2b we study the alternative decays of
$A^0$ and $H^0$ into $t\bar{t}$ final states
for parameter values in Region II.
Here, the minimum value of the lightest stop mass has to be specified and we
take
$m_{\tilde{t}_1}=65\,GeV$.
Even though $A^0\rightarrow b\,\bar{b}$ is also dominant
in Region II for the largest allowed values of $\tan\beta$ in this region, it
has large
QCD backgrounds. The heavy $t\,\bar{t}$ final states, however, are projected in
the
direction of the beam and
can be identified through high-$p_T$ leptons from semileptonic $t$-quark
decays.
Thus the heavy Higgs decays into $t\,\bar{t}$ final states, though they have a
branching fraction smaller than that of the $b\,\bar{b}$ final states for
$\tan\beta\stackM 6$, may in compensation be more manageable from
the experimental point of view.
For these decays we generally select more moderate values for
$\tan\beta$ in order to make them
sufficiently operative.
{}From Fig. 2b we realize that of the two decays into
$t\,\bar{t}$, the most sensitive to SUSY-QCD radiative corrections is that of
the CP-odd Higgs boson.
Here, in contradistinction to the $b$-quark final states,
the main source of the corrections lies in the structure of the
form factors $K_{L,R}$ on
eqs.(\ref{eq:OLOR})-(\ref{eq:KLKR}) -- the top quark self-energies being
negligible.

Of course, we expect --and we have numerically verified-- that the
SUSY-QCD contributions
drop off upon freely increasing the squark masses.
However, in practice the asymptotic regime begins for fairly large values of
these
masses. For example, in Fig.3 we study the Higgs decays into
$b\,\bar{b}$ as a function of $m_{\tilde{b}_1}$, for various $\tan\beta$.
We see that, for $\tan\beta\stackM 10$, the corrections can reach
several $10\%$ even for $m_{\tilde{b}_1}$ in the few hundred $GeV$ range.

Worth noticing is the asymptotic behaviour of
the correction (\ref{eq:deltag}) versus the gluino mass for the various Higgs
decays.
As shown in Figs.4a-4d, it takes a long time, so to speak, for the gluino
to decouple. Corrections of
$\sim 50-60\%$ for $A^0,h^0,H^0\rightarrow b\,\bar{b}$
are obtained at high $\tan\beta$ from a mass value
$m_{\tilde{g}}\simeq 150\,GeV$ all the way out
to $1\,TeV$ -- hence far beyond the present phenomenological bounds.
In Figs.4c-4d we can also assess
 the dependence of $A^0,H^0\rightarrow t\,\bar{t}$
on $m_{\tilde{t}_1}$, for fixed
$m_{\tilde{b}_1}=150\,GeV$ and a moderate value of $\tan\beta$;
and we see that even for
stop masses as heavy as $100\,GeV$ the corrections
are longly sustained (above $10\%$)
for practically any value of $m_{\tilde{g}}$ beyond the
threshold singularities associated to points satisfying
$m_{\tilde{g}}+m_{\tilde{t}_1}\simeq m_t$.
For gluino masses below these points, the
corrections to $A^0\rightarrow t\,\bar{t}$ can be much larger.

{}From Figs.5a and 5b we read off the dependence of
the SUSY-QCD corrections on $M_{A^0}$
for different values of $\tan\beta$, and they
are compared with the ordinary QCD corrections.
We remind the reader that the QCD
corrections to $\Phi^i\rightarrow q\,\bar{q}$ can be very large for
light quarks\,\cite{Braaten,Drees}. As in the decay of
the charged Higgs\,\cite{Ricard},
this is due to the appearance of a logarithmic term
carrying a quark mass singularity,
$\sim\log\,(M_{\Phi^i}/ m_q)$, which stems from the anomalous dimension of
the $\bar{q}\,q$ and $\bar{q}\,\gamma_5\,q$ operators.
The complete one-loop (and renormalization group improved) formulae
read as follows\footnote{This equation is equivalent to eq.(4.5) of
Ref.\cite{Drees},
except that we have corrected a missing factor of $2$ in the last logarithm.}
($b=\frac{33-2n_f}{6\pi}$):
\beq
\Gamma^i(q)=\Gamma_0^i(q)\,\left[1-b\,\alpha_s(M_{\Phi^i})
\,\log{M_{\Phi^i}\over 2\,m_q}\right]^{4/b\,\pi}\,
\left\{1+{C_F\,\alpha_s(M_{\Phi^i})\over\pi}(\Delta_{\Phi^i}+3\,
\log{M_{\Phi^i}\over 2\,m_q})\right\}\,,
\label{eq:deltaQCD}
\eeq
where the complicated functions $\Delta_{\Phi^i}$\, are given by eqs.(3.7) and
(2.26) of
Ref.\cite{Drees} for $i=1$ and $i=2,3$ respectively\footnote{We have also
corrected
a missing factor of $2$ in the third term on the RHS of eq.(2.27) of
Ref.\cite{Drees}.}.
{}From eqs.(\ref{eq:deltaQCD}) and (\ref{eq:treelevel})
the standard QCD corrections
$\delta_g^i=(\Gamma^i_{QCD}(q)-\Gamma_0^i)/\Gamma_0^i$ to the various MSSM
neutral Higgs
decays can be computed and are included
in Figs.5a and 5b, where they can be compared
with the SUSY-QCD effects ($\delta^i_{\tilde{g}}$).

{}From Fig.5a we see that
the decays $\Phi^i\rightarrow b\,\bar{b}$
receive negative standard QCD corrections around
$30-45\%$ (Notice that we have plotted $-\delta_g^i$ in Fig.5a.).
For $M_{A^0}\stackrel{\scriptstyle >}{{ }_{\sim}} 100\,GeV$,
the standard QCD correction to $h^0\rightarrow b\,\bar{b}$
remains saturated at about $-30\%$ since
the mass $M_{h^0}$ also saturates at its maximum value, whereas the modes
$A^0, H^0\rightarrow b\,\bar{b}$ obtain slowly increasing negative corrections.
In contrast, $A^0\rightarrow t\,\bar{t}$ and $H^0\rightarrow  t\,\bar{t}$,
receive positive standard QCD corrections rapidly varying with
the Higgs mass (Cf. Fig.5b)
Comparison with our Figs.2-5 clearly
shows that in many cases the supersymmetric effects are important since
they can be of the same order as the standard QCD corrections.

Overall, large SUSY-QCD quantum corrections are expected in
the hadronic widths of the neutral Higgs bosons of the MSSM.  They should be
measurable, though with different techniques,
both in $e^+\,e^-$ and in hadron machines.
These supersymmetric effects can not only be
comparable to the ordinary QCD corrections, but even dominant in some cases.
Since they can have either sign, the net QCD correction would be
found either much ``larger'' than expected,  perhaps ``missing''
or even with the ``wrong'' sign; in any case, it should be revealing to hint at
the
SUSY nature of these higgses.
Furthermore, we have found that, contrary to the situation with
SUSY corrections on gauge
boson observables, these effects decouple very slowly, especially with the
gluino mass.
Therefore, if SUSY is there, these corrections should also be
there, and cannot be missed for a wide range of sparticle masses.
However,  Region II is out of reach of
LEP 200, and even though part of the Higgs spectrum characterizing Region I
is within its discovery range a complete experimental account of
the MSSM Higgs sector
will not be possible at LEP 200.
In this respect, we have put forward the convenience of trying to
see the kind of effects studied here in the large hadron machines,
perhaps before an
$e^+\,e^-$ supercollider be at work. In fact, the potentially large size
of these effects  indicates that they ought to be included in any serious
analysis of supersymmetric Higgs production
processes in hadron colliders.
The combined information
on branching ratios (from $e^+\,e^-$ ) and on cross-sections
(from the Tevatron and/or at the LHC)
should be very useful to pin down the nature of the observed effects.
A more complete study should include the electroweak SUSY effects (in
particular,
the effect from the one-loop Higgs mass relations), but as already mentioned in
the companion paper\,\cite{Ricard} these are expected not to drastically alter
the SUSY-QCD picture presented here. Our general conclusion is
that quantum corrections on Higgs physics may be the clue to ``virtual''
Supersymmetry.

%\newpage
%\vspace{0.5cm}

{\bf Acknowledgements}:

\noindent
J.S. is indebted to F. Palla for
discussions.
This work has been partially supported by CICYT
under project No. AEN95-0882.

\baselineskip=5.6mm
%\newpage
%\vspace{0.5cm}

\vspace{0.1cm}
\hyphenation{ty-pi-cal}
%\newpage
\vspace{1cm}
\begin{center}
\begin{Large}
{\bf Figure Captions}
\end{Large}
\end{center}
\begin{itemize}
\item{\bf Fig.1} {\bf (a)} Typical mechanisms for neutral Higgs
production at hadron colliders;
{\bf (b)} SUSY-QCD Feynman diagrams, up to one-loop level,
correcting the partial widths of $A^0, h^0, H^0\rightarrow\,q\,\bar{q}$.

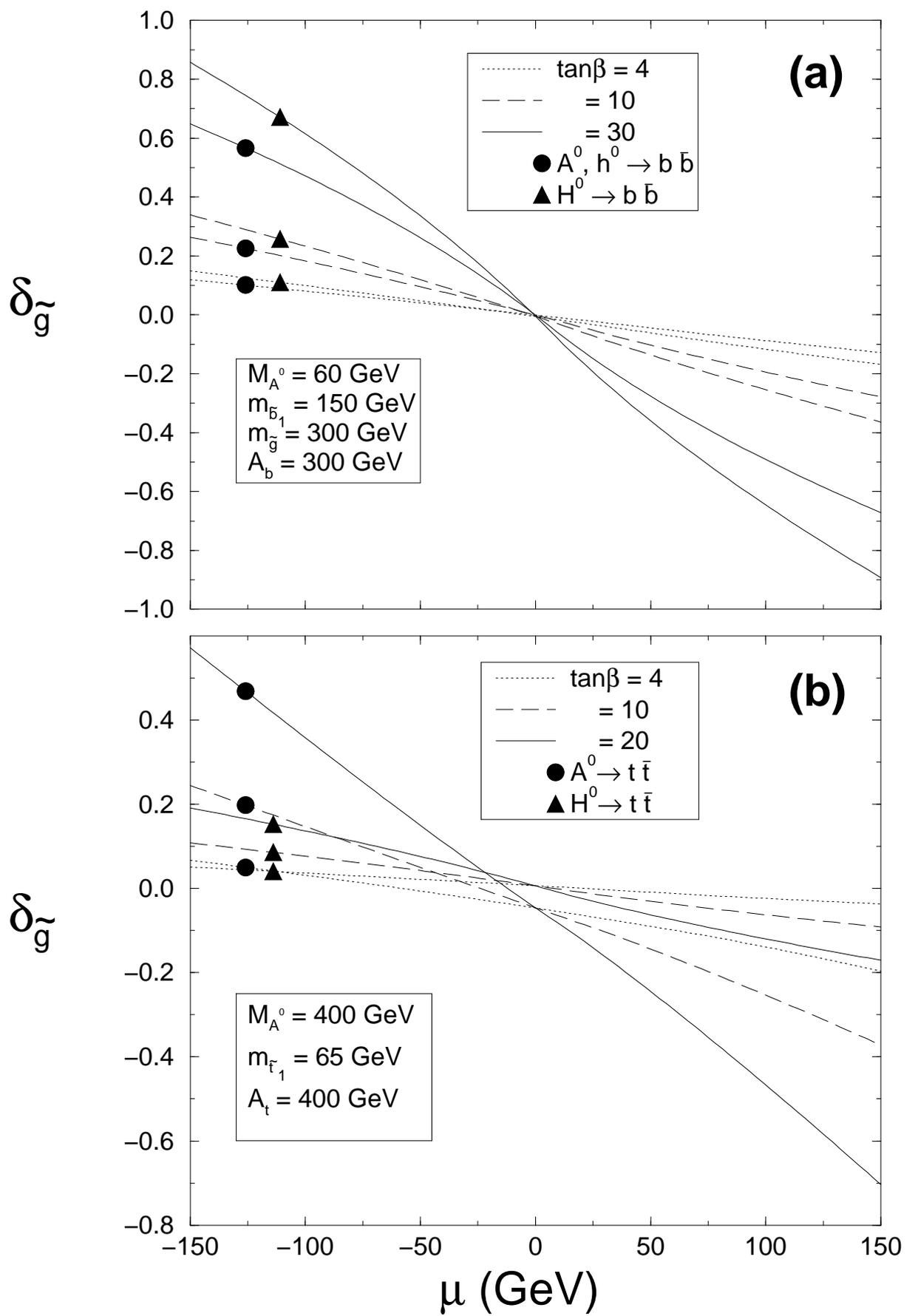
\item{\bf Fig.2} {\bf (a)} Dependence of the relative SUSY-QCD corrections\,
$\delta_{\tilde{g}}^i(q)$ -- eq.(\ref{eq:deltag}) -- for $i=1,2,3$ and $q=b$,
upon the supersymmetric Higgs mass mixing term, $\mu$, for light
$M_{A^0}=60\,GeV$
and given values of the other parameters
(the scale of the abscissa is common to (b) below); {\bf (b)} As before but for
$\delta_{\tilde{g}}^{1,3}(t)$ and heavy $M_{A^0}=400\,GeV$, for
fixed $m_{\tilde{t}_1}=65\,GeV$.

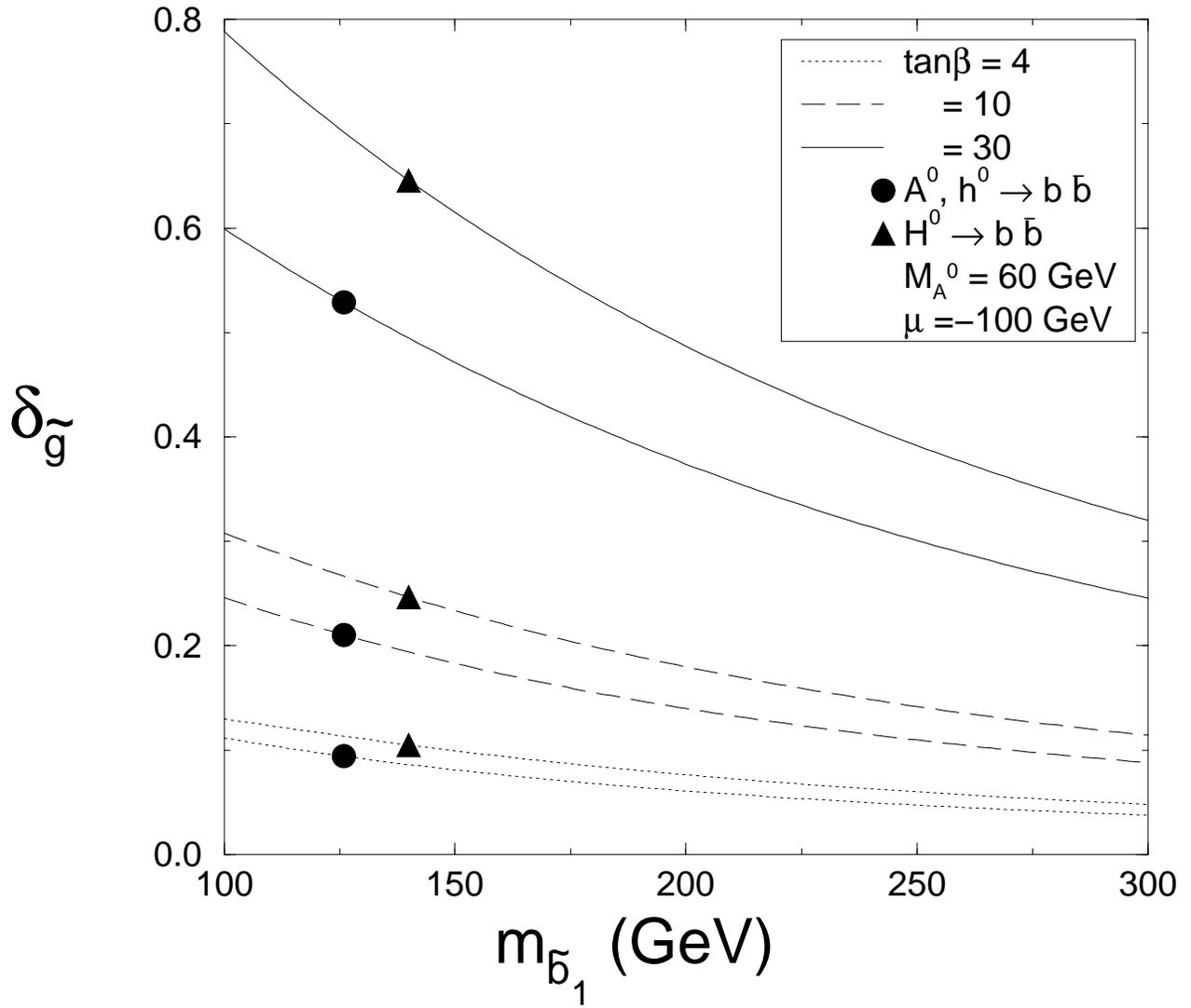
\item{\bf Fig.3}  $\delta_{\tilde{g}}^{1,2,3}(b)$
as a function of the lightest sbottom mass $m_{\tilde{b}_1}$
for various $\tan\beta$ and fixed $\mu$ and $M_{A^0}$.
The other fixed parameters are as in Fig.2.

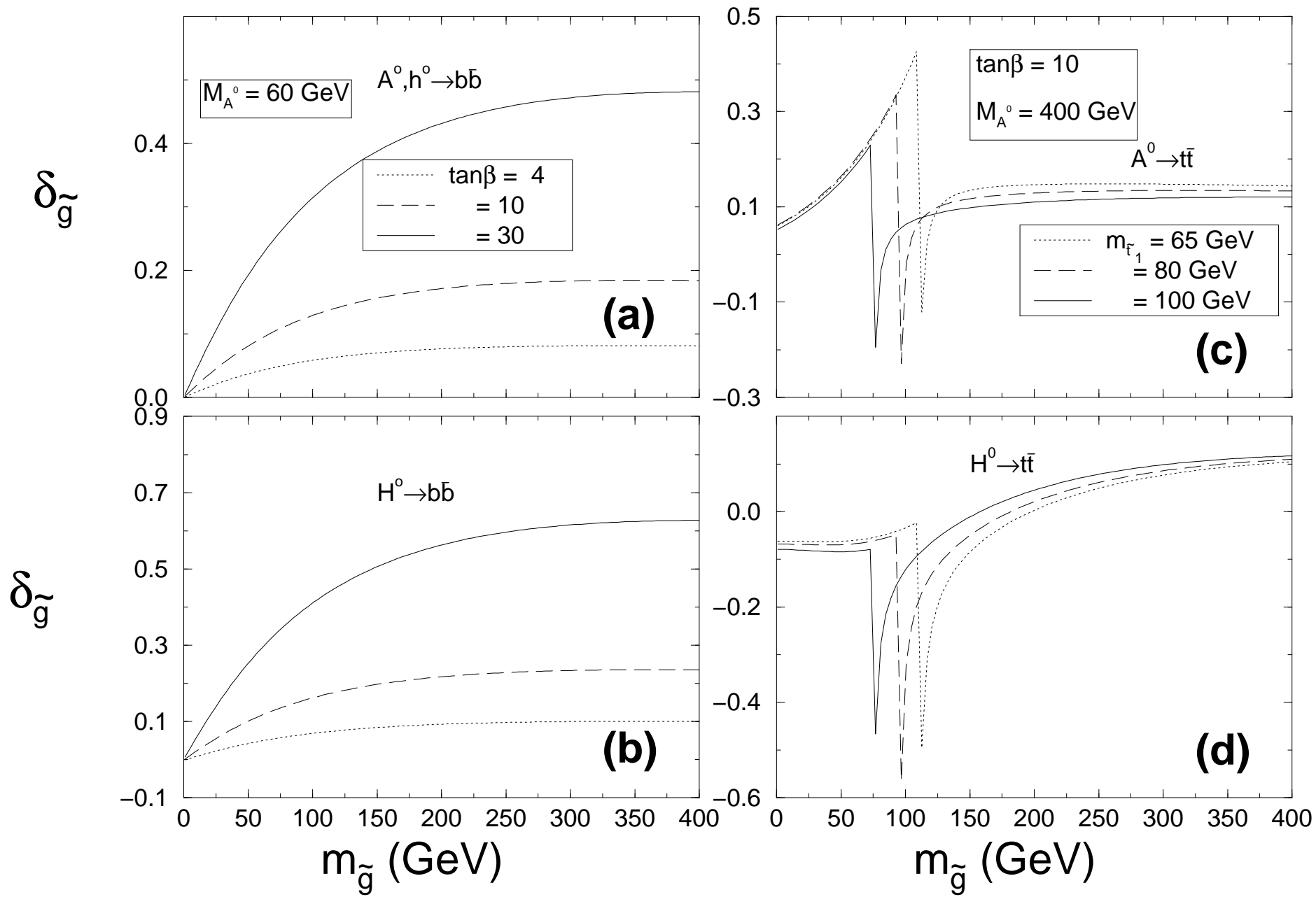
\item{\bf Fig.4} {\bf (a)} Evolution of $\delta_{\tilde{g}}^{1}(b)\simeq
\delta_{\tilde{g}}^{2}(b)$ (almost indistinguishable)
in terms of the gluino mass for various $\tan\beta$.
$M_{A^0}$ is chosen light and the remaining inputs as in
Fig.2 (the scale of the abscissa is common to (b) below);
{\bf (b)} As before but for $\delta_{\tilde{g}}^{3}(b)$; {\bf (c)}
$\delta_{\tilde{g}}^{1}(t)$ versus the gluino mass at fixed $\tan\beta$
and for various $m_{\tilde{t}_1}$
(the scale of the abscissa is common to (d) below);
{\bf (d)} As before but for $\delta_{\tilde{g}}^{3}(t)$.

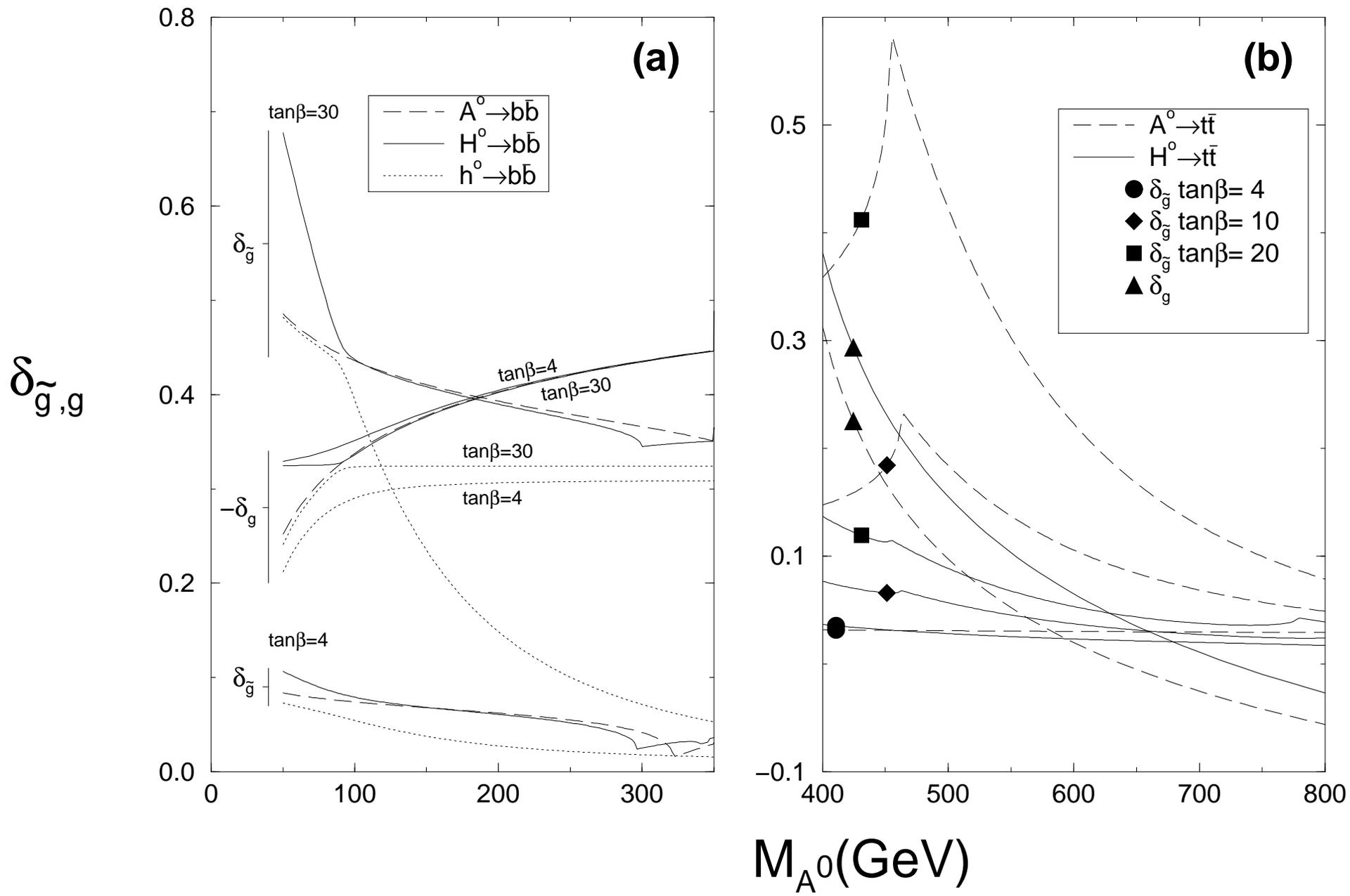
\item{\bf Fig.5} {\bf (a)}  $\delta_{\tilde{g}}^{1,2,3}(b)$ for $\tan\beta=30$
(upper-born curves) and $\tan\beta=4$ (lower-born curves)
as a function of $M_{A^0}$ and the rest of the parameters as in Fig.2.
The middle-born curves stand for the corresponding standard QCD
corrections, $\delta_g^i$, to the three decay processes. The latters being
negative, we plot $-\delta_g^i$ to ease comparison
with the SUSY-QCD corrections at $\mu<0$;
{\bf (b)} As before but for $\delta_{\tilde{g}}^{1,3}(t)$ and the
range of $M_{A^0}$ selected deep into Region II and three values
of $\tan\beta$. The scale of the ordinate is common to (a).

\end{itemize}

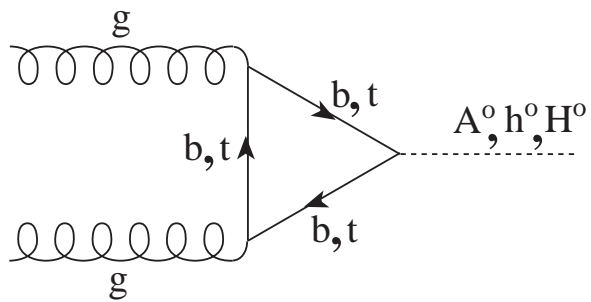
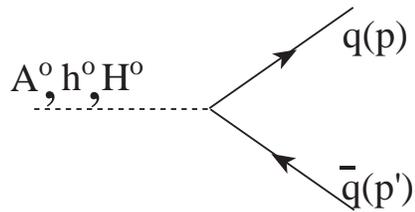
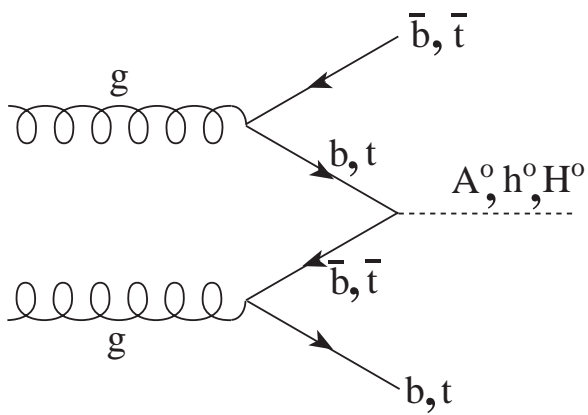
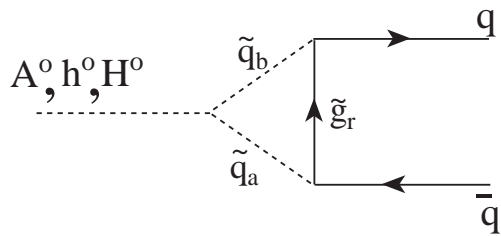
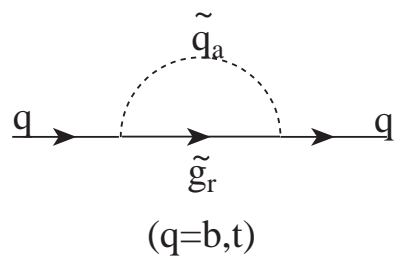
\end{document}